\documentstyle[multicol,preprint,aps,graphicx,psfig]{revtex}

\begin{document}

\title{The self-organized critical forest-fire model on large scales}

\author{Klaus Schenk$^1$, Barbara Drossel$^2$, Franz
Schwabl$^1$
}
\address{${}^1$ Physik-Department der Technischen Universit\"at M\"unchen, James
Franck Stra\ss e, D-85747 Garching}
\address{${}^2$ Physics Department, Raymond and Beverley Sackler Faculty of
Exact Sciences, Tel Aviv University, Tel Aviv 69978, Israel}
\date{\today}
\maketitle

\begin{abstract}
  We discuss the scaling behavior of the self-organized critical
  forest-fire model on large length scales. As indicated in earlier
  publications, the forest-fire model does not show conventional
  critical scaling, but has two qualitatively different types of fires
  that superimpose to give the effective exponents typically measured
  in simulations. We show that this explains not only why the exponent
  characterizing the fire-size distribution changes with increasing
  correlation length, but allows also to predict its asymptotic value.
  We support our arguments by computer simulations of a coarse-grained
  model, by scaling arguments and by analyzing states that are created
  artificially by superimposing the two types of fires.  
\end{abstract}

% \begin{multicols}{2}
\narrowtext
\section{Introduction}
During the past years, systems which exhibit self--organized criticality (SOC)
have attracted much attention, since they might explain part of the abundance
of fractal structures in nature \cite{bak87}. Their common features are slow
driving or energy input and rare dissipation events which are instantaneous on
the time scale of driving.  In the stationary state, the size distribution of
dissipation events obeys a power law, irrespective of initial conditions and
without the need to fine-tune parameters. Examples for such systems are the
sandpile model \cite{bak87}, the self-organized critical forest-fire model (FFM)
\cite{dro92,hen93,gra93,cla94}, the earthquake model by Olami, Feder, and
Christensen \cite{ola92}, and the Bak-Sneppen model \cite{sne93,PMB}.

The study of SOC models is usually based on the assumption that the size
distribution $n(s)$ of dissipation events (avalanches, fires, earthquakes)
shows the scaling behavior familiar from equilibrium critical systems,
\begin{equation}
n(s) \simeq s^{-\tau}{\cal C}(s/s_{\small{\rm max}})\, , \label{eq_scaling}
\end{equation}
with a cutoff function ${\cal C}$ that is constant for small arguments
and decays exponentially fast when the argument is considerably larger
than 1. The cutoff cluster size $s_{\small{\rm max}}$ is related to
the correlation length $\xi$ via $s_{\small{\rm max}} \sim \xi^D$,
with $D$ being the fractal dimension of the dissipation events. (If
the cutoff is set by the system size $L$, $\xi$ must be replaced with
$L$.) This holds indeed for some self organized critical systems, like
the Bak-Sneppen model, but it has been known for some time that it
does not hold for the two-dimensional Abelian sandpile model
\cite{teb98}. Very recent work has shown that this violation of simple
scaling in the sandpile model is due to the existence of multiple
waves of topplings, and some features of the correct scaling behavior
have been worked out \cite{ste99,dro00}. Violation of finite-size
scaling is also seen in the above-mentioned earthquake model
\cite{pac01}.

During recent years, evidence has accumulated that the two-dimensional
SOC forest-fire model does not show simple scaling either. Instead,
there are more than one diverging length scale \cite{hon96}, the
behavior of the model for tree densities just above the critical
density is completely different from that of conventional critical
systems \cite{cla95,cla95b}, and finite-size scaling is violated
\cite{sch00}. A scaling collapse based on Eq.~(\ref{eq_scaling}) gives
a good overlap of the tails of the distribution, but not so much of
the first part, where the slope (i.e., the exponent $\tau$) seems to
increase slightly with increasing correlation length (see, e.g.,
\cite{gra93}, and
Figure~\ref{fig_original_ffm_ns_superpos_and_tendency} below).  We have
suggested \cite{sch00} that all these features are due to the fact
that there are two qualitatively different types of fires in the
system: smaller, fractal fires that occur in regions of low tree
density and burn a tree cluster that resembles a percolation cluster,
and larger compact fires that burn a patch of a tree density above the
percolation threshold.

It is the purpose of this paper to show how these two types of fires add up to
give the distributions typically seen in computer simulations, and to derive
the asymptotic properties of the fire size distribution in the limit of very
large correlation length. In particular, we will derive the asymptotic value of
the exponent of the fire size distribution, toward which it should
converge for sufficiently large correlation length. We support our arguments by
three different types of evidence that all lead to the same conclusions: (i)
Scaling arguments based on the superposition of the mentioned two types of
fires;
%and on the finding that the mean tree density approaches its critical
%value in a manner different from equilibrium critical systems; 
(ii) The fire
size distributions resulting from the (artificial) superposition of the fires
of patches of different tree density and different  size; 
(iii) Computer simulations of a
coarse-grained model that allow us to study numerically systems with much
larger correlation length than has been possible so far.  The outline of the
paper is as follows: In the next section, we will derive scaling laws and
analytical expressions for the fire size distribution resulting from the
assumption that the forest-fire model is composed of patches of different size
and different tree densities. Then, in section~\ref{chap_superpose} we will show
numerical data that result from the superposition of fires from artificially
generated patches of different sizes and densities. In section~\ref{chap_fleckmodell}, 
we will present and study a coarse-grained forest-fire model
where each lattice site stands for a group of several sites in the original
model. Finally, we will discuss our findings.

\section{Scaling properties of the SOC forest-fire model}
\label{chap_scaling_laws}

\subsection{The definition of the model}
The SOC forest-fire model is usually studied on a square lattice with $L^2$
sites. Each site is either occupied by a tree, or it is empty. At each time
step, all sites are updated in parallel according to the following rules: (i)
An empty sites becomes occupied by a tree with probability $p$; (ii) A tree is
struck by lightning with probability $f$. This tree and the whole cluster of
trees connected to it (by nearest-neighbor coupling) burn down and become
empty sites.

\subsection{Two types of fires determine the dynamics of the SOC forest fire model}
As long as $p$ and $f$ are so small that fires do not interfere with
each other or with tree growth, the stationary behavior of the model
depends only on the ratio $f/p$, but not on the two parameters
separately.  After some time, the system reaches a stationary state
with a mean tree density $\rho$ and a mean fire size $\bar s$. A
snapshot of the $2$d system in the stationary state is shown in
Figure~\ref{fig_snapshot}. One can see that it consists of patches of
different tree density and different sizes. Some of the patches have a
high tree density, and if they are struck by lightning, the entire
patch burns down, with only few trees being left. After the fire, the
tree density $\rho_{\rm{\small patch}}(t)$ of the patch grows
again according to
\begin{equation}
\dot \rho_{\rm{\small patch}}(t)  = p(1 - \rho_{\rm{\small patch}}(t))
\label{eq_tree_growth}
\end{equation}
until it is hit by the next lightning stroke.  This mechanism of growth and
burning down of forest clusters produces the patchy structure seen in
Figure~\ref{fig_snapshot}, which is characterized by the following
properties.

\begin{itemize}
\item The patches are almost homogenously covered with trees.  This is
because a fire that burns a patch usually leaves only a few trees
behind. (We found that the local tree density within a patch
immediately after a fire is typically $0.078$ for the $2$d square
lattice and $0.062$ for the $2$d triangular lattice). Thus the random
tree growth leads to a uniform tree density within the patch.  The
size distribution of tree clusters within a patch is therefore similar
to the size distribution of clusters in a percolation system of the
same density. (For an introduction to percolation theory, see
\cite{sta92}.) The fact that several patches contain larger dense tree
clusters indicates that some fires leave behind small clusters
that can become seeds of new patches. This process of birth and death
of patches, which happens on a slower time scale, does not affect our
main argument.
\item It can be assumed that the distribution of patch sizes is 
independent of their tree density, because a fire in most cases
hardly changes the size of a patch. Thus, the size of most patches
is the same for high and low tree density.
\item Some patches have a tree density above the percolation
threshold. These patches contain a spanning cluster that is compact,
i.e., has the fractal dimension 2.  When such a patch is struck by
lightning, a compact fire occurs that burns the spanning cluster.  If
lightning strikes a patch of low tree density (below the percolation
threshold), only a small, fractal cluster of trees burns down, and
only part of the patch is affected by the fire.  
If the mean fire size is large (i.e.\ if $f/p$
is small), most trees burn down during large fires, and most of the
empty sites are created during the large compact fires, resulting 
in the above-mentioned low local tree density immediately after a fire. 
\item 
The size of the largest patch diverges for $f/p \to 0$, suggesting
that the system is close to a critical point and can be characterized
by power laws. Several such power laws will be mentioned further below.
\end{itemize} 

The two mentioned types of fires add up to give the distributions
typically seen in computer simulations, and explain the unconventional
behavior of the forest fire model mentioned in the previous section.
The left part of the fire size distribution of the forest fire model
(see Figure~\ref{fig_original_ffm_ns_superpos_and_tendency}b) is
mainly due to fires burning fractal percolation clusters, and the
cutoff part is due to large compact fires. In contrast, in
conventional critical systems the power-law part and the cutoff part
are due to the same type of critical fluctuations.  Since the two
parts of the distribution become clearly separated only for very large
$\xi$, as we will show below,  the asymptotic exponent of the
fire-size distribution is not visible in present-day computer
simulations.

\subsection{The scaling behavior}
In the stationary state, the mean number of growing trees must equal the mean
number of trees burning down, leading to \cite{dro92}
\begin{equation}
\bar s = \frac{p(1-\rho)}{f \rho}.\label{eq_sbar}
\end{equation}
$\bar s$ diverges according to a power law in the limit $f/p \to 0$,
implying that the size $s_{\small{\rm max}}$ of the largest fires also
diverges, and with it the correlation length $\xi$, which we define to
be the radius of the largest fires. The cutoff fire size
$s_{\small{\rm max}}$ can be expected to scale as
\begin{equation}
s_{\small{\rm max}} \sim (f/p)^{-\lambda},
\label{eq_smax_lambda_scaling_formel}
\end{equation}
with an exponent $\lambda$ \cite{hen93,gra93,cla94}, which has a value
close to 1.1. This leads to $\xi \sim (f/p)^{-\lambda/2}$ since the
fractal dimension of the large fires is $2$.  (In contrast, earlier
work was based on the assumption that large and small fires have the
same fractal dimension, which was found to be $D \simeq 1.96$
\cite{cla94}, with some authors not ruling out the value $D=2$
\cite{gra93,hen93}.)

\begin{figure}
\begin{center}
\includegraphics[angle=0, width=\textwidth]{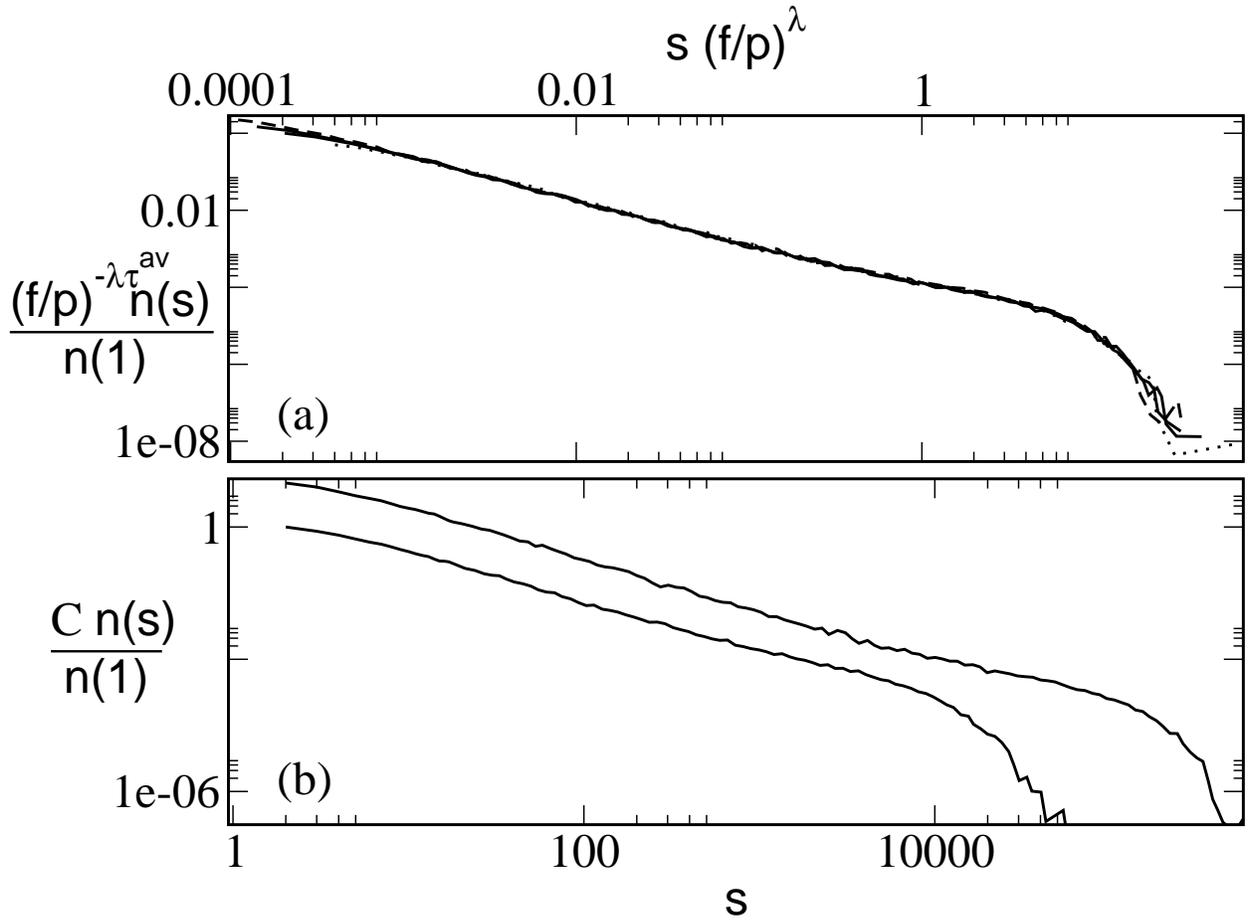}
\caption{(a) Superposition of the fire size distributions $n(s)$ of
  the SOC forest fire model for $L=1300$ and $f/p= 0.000118$ (dashed
  line), $0.000169$ (solid line), $0.000236$ (solid line) and
  $0.000394$ (dotted line). In order to make the curves collapse, the
  vertical axis had to be scaled with $(f/p)^{-\lambda \tau^{\rm\small
      av}}$, using the effective exponent $\tau^{\rm\small av}=1.14$, which
  can be interpreted as an average exponent over a certain range of
  $s$ values. (b) Fire size distributions $n(s)$  for
  $L=1300$ and $f/p=0.0001183$ (upper curve) and
  $f/p=0.001183$ (lower curve). The upper curve was shifted vertically by a factor $C=10$, in order to make the shapes of the two curves better visible.}
\label{fig_original_ffm_ns_superpos_and_tendency}
\end{center}
\end{figure}

Let $s$ be the number of trees burnt during a fire, and $n(s)$ the size
distribution of  fires in the system, normalized such that $\int n(s) ds =
\rho$. Since each tree is struck by lightning with the same probability, the
size distribution of tree clusters is proportional to $n(s)/s$. We write
\begin{equation}
n(s) = n_1(s) + n_2(s),
\label{eq_2feuer_scaling_hauptgleichung}
\end{equation}
with $n_1(s)$ being the contribution from the smaller, fractal fires, and
$n_2(s)$ being the contribution from the compact fires that burn an entire
patch. If there was no qualitative difference between the two types of fires,
the scaling law Eq.~(\ref{eq_scaling}) would hold, from which one could derive the
scaling relation 
\begin{equation}
 \lambda = \frac{1}{2-\tau},
\label{eq_lambdatau_skalenrelation}
\end{equation}
(This is obtained from the condition $\bar s = \int_1^\infty s n(s)
ds/\rho$ and is given in many earlier publications.)  However, since
we have two different types of fires, we cannot expect this relation
to hold.  Figure~\ref{fig_original_ffm_ns_superpos_and_tendency}b
shows the size distribution of fires for different values of $f/p$.
(The system size $L$ was chosen large enough to avoid finite-size
effects.)  One can see that $n(s)$ becomes steeper with decreasing
$f/p$, and one can expect the slope to grow further until it reaches
some limit value.  The steepest slope occurring in this figure has the
absolute value $\tau\approx 1.3$,
which must be a lower bound to the asymptotic value of $\tau$. (The
value $\tau \simeq 1.14$ given in many earlier publications was
obtained by taking some average value along $n(s)$, which was
less than its steepest slope.) A scaling collapse (see
Figure~\ref{fig_original_ffm_ns_superpos_and_tendency}a) shows that
the cutoff parts of the curves superimpose nicely, allowing to derive
a value $\lambda \simeq 1.1$, which does not fit together with the
scaling relation Eq.~(\ref{eq_lambdatau_skalenrelation}). The same
result for $\lambda$ was obtained by Pastor-Satorras and Vespignani
\cite{pas00} using a moment analysis, confirming that the cutoff shows
simple scaling behavior.

Next, let us discuss the properties of $n_1(s)$. As $f/p$ decreases, the first
part of $n(s)$ does not change any more. This indicates that $n_1(s)$ reaches
an asymptotic form $n_1^*(s)$ as $f/p \to 0$, with a cutoff that depends on $f/p$. We therefore write
\begin{equation}
n_1(s) = n_1^* (s)  {\mathcal{C}_{\rm 1}}(s/s_{\rm{\small max,fractal}})=
n_1^* (s)  {\mathcal{C}_{\rm 1}}(s(f/p)^{\lambda_1}),
\label{eq_n1s}
\end{equation}
introducing the cutoff function ${\mathcal{C}_{\rm 1}}(s/s_{\rm{\small
    max,fractal}})$ for the distribution of the fractal fires and
assuming that the maximum fractal fire size $s_{\rm{\small
    max,fractal}}$ scales with an exponent $\lambda_1$.  For
sufficiently large $s$, $n_1^*(s)$ will reach an asymptotic power law
with the "true" exponent $\tau$. We can estimate the value of $\tau$
from the following argument: The large fractal fires stem from the
percolation clusters in those patches that have a tree density close
to the percolation threshold $\rho_{\small{\rm perc}}$. Thus the
probability density of finding a cluster of size $s$ is proportional to the
probability that $\rho_{\mbox{\small patch}}$ of a large patch is
large enough that percolation clusters of size $s$ exist, multiplied by the
probability density to find a cluster of size $s$ in a system at the
percolation threshold.  The fire size distribution is proportional to
$s$ times the cluster distribution, as we mentioned above.  The
probability to find a cluster of size $s$ in a percolation system is
determined by the size distribution of percolation clusters:
\begin{equation}
n_{\small{\rm perc}}(s) \simeq s^{-\tau_{\small{\rm perc}}} 
{\mathcal C}_{\small{\rm perc}}(s/s_{\rm{\small max,perc}})
\label{eq_percolation_n_s}
\end{equation}
with 
\begin{equation}
s_{\rm{\small max,perc}}\propto 
(\rho_{\small{\rm perc}}-\rho_{\small{\rm\small patch}})^{-\sigma_{\small{\rm perc}}}
\label{eq_sigma_perc_def}
\end{equation}
and $\sigma_{\small{\rm perc}}=91/36\approx 2.528$ and
$\tau_{\small{\rm perc}}=187/91\approx 2.055$ \cite{sta92}.  In these
patches close to the percolation threshold, the tree density increases
with time approximately as $\dot \rho_{\small{\rm patch}} =
p(1-\rho_{\small{\rm perc}})$.  Therefore, the probability that the
density is within a distance $\rho_{\small{\rm
    perc}}-\rho_{\small{\rm patch}}$ of the percolation threshold is
proportional to $\rho_{\small{\rm perc}}-\rho_{\small{\rm patch}}$,
and the probability that a patch has tree clusters larger than $s$ is
proportional to $s^{-1/\sigma_{perc}}$ (see
Eq.~(\ref{eq_sigma_perc_def})).  The probability that a fire of
size $s$ occurs is consequently proportional to
$$ s^{1-\tau_{perc}-1/\sigma_{perc}} \simeq s^{-1.45}.$$
This means 
$$\tau \simeq 1.45.$$

Next, let us estimate the cutoff exponent $\lambda_1$. As we have seen
above, the radius $\xi$ of the largest patches is proportional to
$(f/p)^{-\lambda/2}$. The size of the largest fractal tree clusters is
therefore proportional to $(f/p)^{-D_{\small{\rm perc}}\lambda/2}$,
with the fractal dimension $D_{\small{\rm perc}}$ of percolation
clusters $\simeq 1.56$, implying $\lambda_1 \simeq 0.86$.  Beyond the
cutoff size for fractal clusters, proportional to
$(f/p)^{-\lambda_1}$, the size distribution of fires must be dominated
by the compact fires and therefore by the size distribution of
patches. Since $\lambda_1 < \lambda$, the "bump" (which is dominated
by the compact fires) should span a larger fraction of the fire size
distribution for smaller $f/p$. Figure~\ref{fig_original_ffm_ns_superpos_and_tendency}b
shows that this is indeed the case.

Assuming that the patch size distribution $n_{\rm{\small patch}}(s)$
scales, too, we suggest a scaling form
\begin{equation}
n_{\rm{\small patch}}(s) \simeq s_{\rm{\small max}}^{b-2} s^{-b} {\mathcal{C}_{\rm{\small 2}}}(s/s_{\rm{\small max}}),
\label{eq_patchdistribution}
\end{equation}
with $s$ being the number of sites in a patch, and $s_{\rm{\small
max}} \propto (f/p)^{-\lambda}$ being the area of the largest patch.
Since most of the system is covered by large patches, $b$ must be
smaller than 2, requiring the factor $s_{\small\rm{
max}}^{b-2}$ in eq.~(\ref{eq_patchdistribution}) in order to normalize
$\int sn_{\rm{\small patch}}(s)ds$.  The size distribution $n_2(s)$
for the compact clusters depends on $n_{\rm{\small patch}}(s)$, but
the relation between the two is non trivial. The reason is the
following: Assume that a patch is struck by lightning always when its
density is so far above the percolation threshold that it burns down
completely. In this case, the size distribution of the large fires
would be proportional to $s$ times the size distribution of the
patches. However, in this case patches would never be destroyed. On
the other hand, patches merge from time to time with neighbors, when
the neighbor reaches a density above the percolation threshold before
lightning strikes the patch with the higher density. In order to
obtain a stationary patch size distribution, patches must therefore be
destroyed from time to time. This can only happen if they are hit by
lightning with a non vanishing probability as long as their density is
sufficiently close to the percolation threshold, such that smaller
dense clusters of trees are left behind by the fire that can develop
into small new patches. For this reason the size distribution of the
large fires is different from $s$ times the size distribution of
patches. This will be seen also in the next section.

\subsection{The exponent $\delta$ }
Additional support for the picture that the fire size distribution is the sum
of two qualitatively different contributions comes from the scaling behavior
of the tree density. It has been known for a long time that the tree density
approaches its critical value according to 
$$ (\rho_c -\rho) \sim (f/p)^{1/\delta},$$
with $1/\delta \simeq 0.5$ \cite{hen93,gra93,cla94}. (The most recent
and probably most accurate value is 0.47 \cite{pas00}.)
If the fire size distribution obeyed the scaling law Eq.~(\ref{eq_scaling}), one
would expect $\delta$ to follow from
$$\rho_c-\rho = \int_{s_{\small{\rm max}}}^\infty s^{-\tau} ds$$
Assuming that conventional scaling (see Eq.~(\ref{eq_scaling})) holds and 
using Eq.~(\ref{eq_smax_lambda_scaling_formel}) and~(\ref{eq_lambdatau_skalenrelation}) this leads to 
$$1/\delta =\frac{\tau-1}{2-\tau}.$$
With the apparent value of $\tau$
around 1.14, this would result in a value of $1/\delta$ much smaller
than 0.47. With the asymptotic value $\tau=1.45$ (see below),
$1/\delta$ would have to be much larger than 0.47. It has been pointed
out recently that the observed value of $\delta$ makes corrections to
simple scaling necessary \cite{pas00}, and a second contribution to
$n(s)$ has been suggested, which has a larger exponent $\tau$ but the
same cutoff as the main contribution, and which becomes negligible for
sufficiently small $f/p$ and sufficiently large $s$. In contrast to
these authors, we argue that there occur not merely corrections to
scaling, but that the scaling behavior of the SOC forest-fire model is
fundamentally different from simple scaling. For this reason, there is
no relation between the exponents $\delta$ and $\tau$, since there is no single
exponent $\tau$ describing the entire fire size distribution. Whether there
exists another relation between $\delta$ and the fire size
distributions $n_1$ and $n_2$, we do not know.

\section{Superposition of different types of fires}
\label{chap_superpose}

In order to show that the fire size distribution seen in simulations
can be indeed the result of the superposition of the two mentioned
types of fires, and in order to confirm that the asymptotic value of
$\tau$ is 1.45, we superimposed the cluster size distributions of
2-dimensional lattices that were homogenously covered with trees, and
that had tree densities between $\rho_{\rm\small after \; the \;
  fire}=0.078$ and $\rho_{\rm\small max}=0.625$, with weights derived
from Eq.~(\ref{eq_tree_growth}). This kind of superposition was sugested by
S. Clar in~\cite{cla95b}. The values for $\rho_{\rm\small
  after\; \; the\; fire}$ and $\rho_{\rm\small max}$ were measured for
instance in~\cite{cla95b}.  But in addition to the superposition of the different
tree densities we  also superimposed different lattice
sizes $l$, distributed according to Eq.~(\ref{eq_patchdistribution})
with $l=\sqrt{s}$ and cutoff
$l_{\rm\small{max}}=\xi=\sqrt{s_{\rm\small{max}}}$. The lattices thus
represent patches of different sizes and densities.  In order to find
the value of the exponent $b$, we performed superpositions for 20
different values of $b$ from 0.1 to 2.  The results did not depend
very much on $b$ as long as $b$ was smaller than $1$, suggesting that
the value of $b$ is in the interval (0,1), but not allowing us to fix
it more precisely.  The results are only reproducible when the
statistics are sufficiently good. For this reason, we had to
superimpose $10^4$ or more systems.

\begin{figure}
\begin{center}
\includegraphics[angle=0, width=\textwidth]{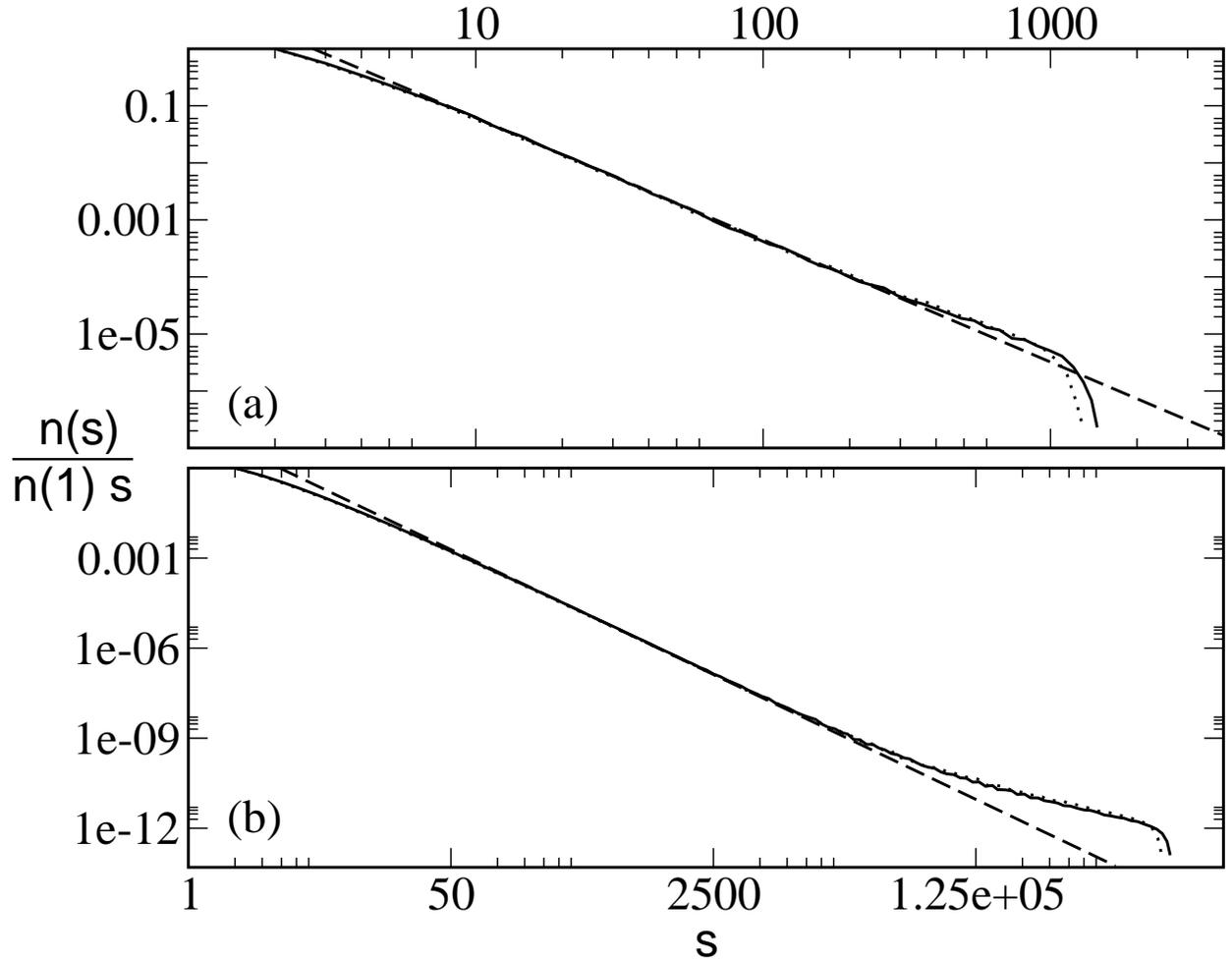}
\caption{(a) Size distribution of tree clusters $n(s)/s n(1)$ resulting from the superposition of lattices with 
$l_{\rm\small{max}}=50$ and $b=0.6$ (solid line: square lattice; dotted line: triangular lattice), and a power law with the exponent $\tau+1$ $=$ $2.14$ (dashed line). (b) The same for  $l_{\rm\small{max}}=2000$ and $b=0.6$, compared to a power law with the exponent $\tau+1$ $=$ $2.45$. The tree densities
we used cover the interval [0.078,0.625] for the square lattice and  [0.062,0.534] for the triangular lattice.}
\label{fig_xi_50_and_2000_superposition}
\end{center}
\end{figure}

The results are shown in
Figure~\ref{fig_xi_50_and_2000_superposition}a and
Figure~\ref{fig_xi_50_and_2000_superposition}b.  Please note that
these figures give cluster size distributions $n(s)/s$ and not fire
size distributions $n(s)$, i.e., the exponents are larger by 1.
Figure~\ref{fig_xi_50_and_2000_superposition}a shows that the apparent
exponent $\tau \approx 1.14$ typically found in simulations of the
$2$d forest fire model is reproduced by the superposition. The smaller
slope for small $s$, and the bump followed by the cutoff, are
reproduced as well. We have performed this superposition also for a
triangular lattice. This lattice is most easily implemented by taking
a square lattice and including next-nearest neighbor couplings along
one of the diagonals in each unit cell. As for the square lattice, the
range of tree densities was obtained from simulations of the SOC
model, and was found to cover the interval [0.062,0.534]. One can see
that the apparent exponent $\tau$ is the same as for the square
lattice, explaining the ``universality'' of this exponent with respect
to a change of the lattice type found earlier \cite{cla94}. (All other
figures shown in this paper are for the square lattice only.)
Figure~\ref{fig_xi_50_and_2000_superposition}b shows that the
distributions of the two cluster types separate for larger correlation
length $\xi$, and that the slope of the part of the curve that stems
from fractal clusters tends to $\tau = 1.45$ as we calculated in
section~\ref{chap_scaling_laws}. A similar effect will be found in the
coarse-grained model discussed in the next section (compare
Figure~\ref{fig_coarse_grained_n_s_comparison_slopes_and_collapse}a).

Our results show also that the size distribution of the largest fires
is related to the size distribution of patches in a nontrivial way, as
mentioned in the previous section. If all large fires did burn
complete patches, the bump of the fire size distribution would have a
slope $-b+1$, which is positive. Since this is not the case, many tree
clusters must be contributing to $n(s)$ that are large but do not
cover the entire patch.  This consideration should hold for any value
of $s_{\small\rm{max}}$. We expect therefore the value of
$\rho_{\rm\small max}$ to decrease slightly with increasing
$s_{\small\rm{max}}$, such that there is always a non vanishing
contribution to $n(s)$ of clusters that are large but do not cover the
entire patch.

We conclude that the superposition of homogeneous patches reproduces
important features of the SOC FFM. It is also an efficient way of
studying the regime of large correlation length, which is not
accessible to direct computer simulations.

\section{A coarse-grained forest-fire model}
\label{chap_fleckmodell}

\subsection{Definition of the model}

In order to be able to study larger systems, we introduced a
coarse-grained model where each site stands for a group of sites in
the original model.  The variable at each site of this coarse-grained
model is the local tree density $\rho_{\rm\small site}$, ranging
continuously from 0 to 1. The rules of our coarse-grained model are the
following: (i) the density at all sites increases per time step by a
small amount $\Delta \rho_{\rm\small site} = p(1-\rho_{\rm\small
  site})$; (ii) Lightning strikes each site with a probability $f$. If
the density of this site is below the percolation threshold
$\rho_{\rm\small perc}=0.59$, nothing happens. If the tree density on
a site struck by lightning is above the percolation threshold
$\rho_{\rm\small perc}$, this site and the entire cluster of sites
above the percolation threshold connected to it burn down. The density
on a site after a fire is a random number between 0 and $r$. The
parameter $r$ takes short-range fluctuations in the density into
account.  The smaller $r$, the smaller the density fluctuations.
Smaller $r$ means consequently that the density at each site is the
average of a larger number of sites in the original model.  We
therefore expect the coarse-grained model for small $r$ to resemble the
original FFM for large $\xi$.

\subsection{Properties of the coarse-grained forest fire model}

\begin{figure}
\begin{center}
\includegraphics[width=\textwidth,angle=90]{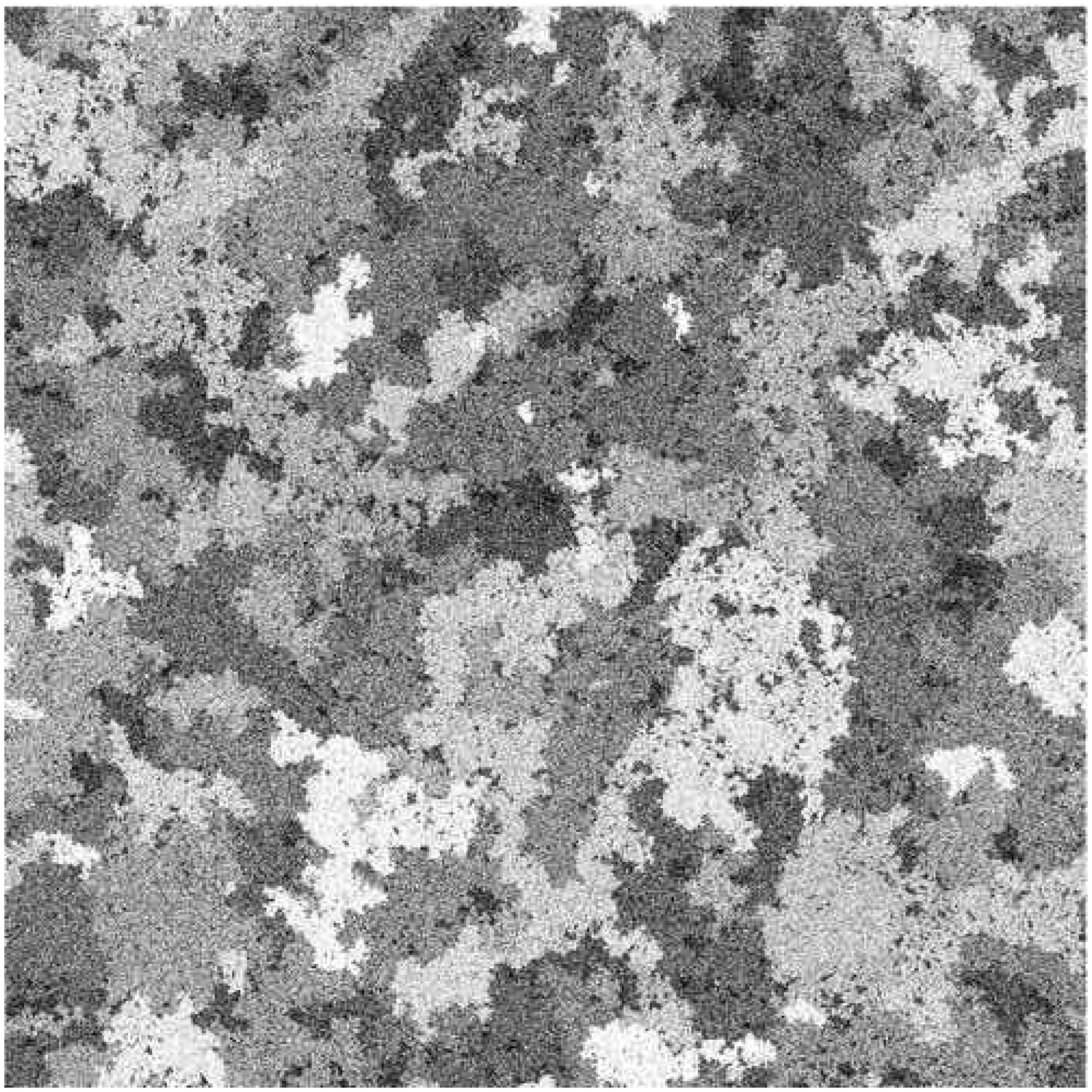}
\caption{Snapshot of the SOC forest-fire model for $\rho \simeq
\rho_c \simeq$ 0.408 and system size $L$=4096. Trees are black and empty sites
are white.} \label{fig_snapshot}
\end{center}
\end{figure}

\begin{figure}
\begin{center}
\includegraphics[width=\textwidth,angle=90]{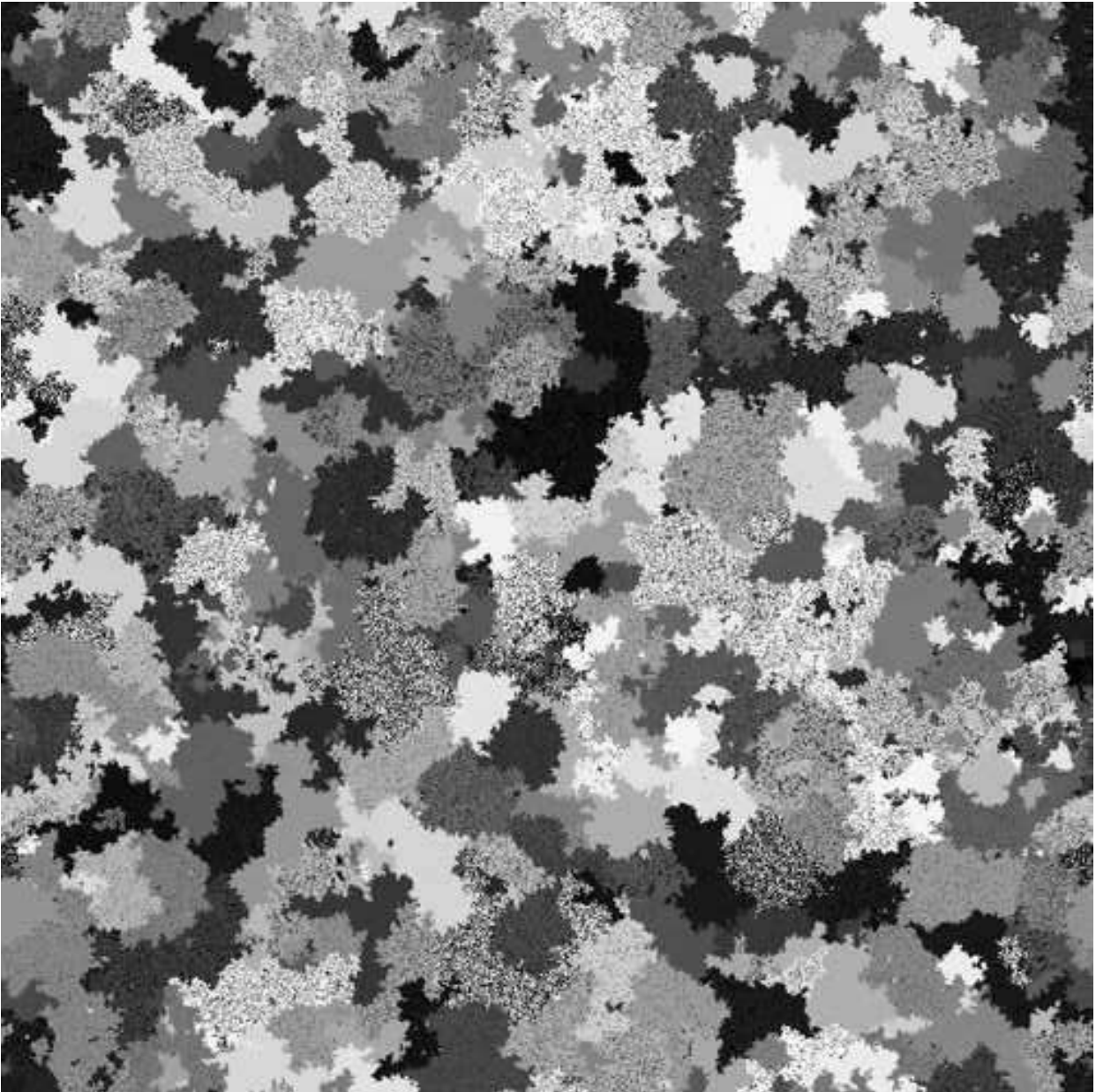}
\caption{Snapshot of the coarse-grained forest-fire model for 
$p=0.01$, $r=0.1$, $f=0.0001$ and $L=1000$. The tree density of a site is 
represented by its grey shade, with larger densities being darker.} 
\label{fig_snapshot_effective_model}
\end{center}
\end{figure}

Although the coarse-grained model is not exactly the same as the
original FFM, it shares many of its features.
Figure~\ref{fig_snapshot_effective_model} shows a snapshot of the
coarse-grained model for $r=0.1$. The figure shows a patchy structure
similar to the one in Figure~\ref{fig_snapshot}. In many patches one
can see sites of two different densities. This indicates that
lightning often strikes a patch before all of its sites (which cover a
range of densities of the width $r$) have a local density above the
percolation threshold, leaving behind some sites with a density just
below the percolation threshold. If this happens several times within
the same patch, one can expect the patch to be destroyed and replaced
by a set of smaller patches. Such processes of birth and death of
patches are not considered further in this paper, but of course they
occur also in the original FFM, as can be seen in
Figure~\ref{fig_snapshot}. As we have mentioned above, creation of new
small patches must occur in order to balance merging and growth of
patches in the stationary state.

\begin{figure}
\begin{center}
\includegraphics[angle=0, width=\textwidth]{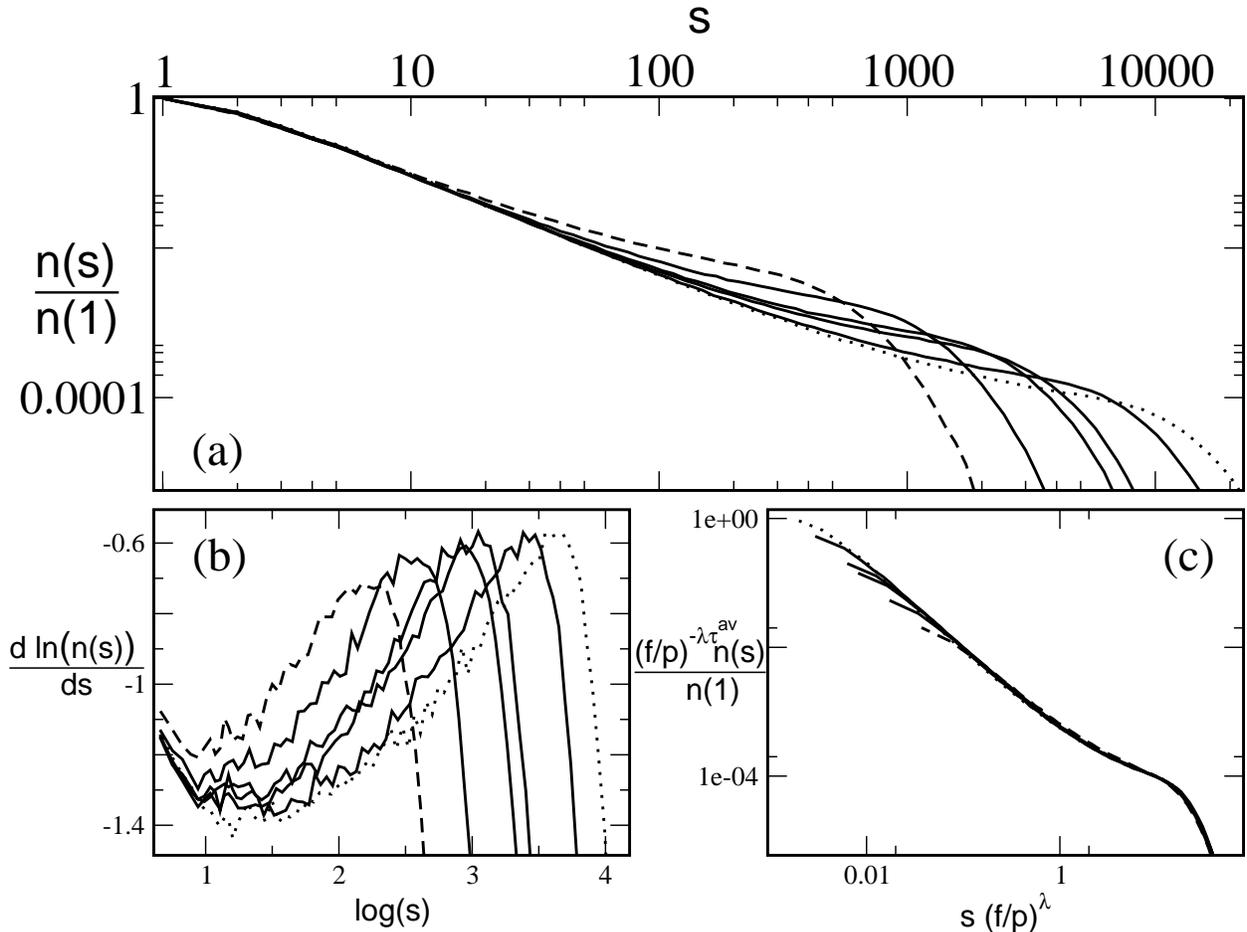}
\caption{(a) Fire size distribution $n(s)/n(1)$ of the coarse-grained forest
fire model for $f/p=$ $0.0031$ (dotted curve), $0.005$,
$0.01$, $0.0125$, $0.025$ and $0.05$ (dashed curve), with
the parameters $r=0.1$ and $L=1000$. (b) Slopes of the curves shown in (a). 
(c) Collapse of the fire size distributions of (a). In order to make the curves collapse, the
  vertical axis had to be scaled with $(f/p)^{-\lambda \tau^{\rm\small
      av}}$, using the effective exponent $\tau^{\rm\small av}=1.25$, which
  can be interpreted as an average exponent over a certain range of
  $s$ values.}
\label{fig_coarse_grained_n_s_comparison_slopes_and_collapse}
\end{center}
\end{figure}

Next, let us consider the size distribution of fires in the coarse-grained
model.  Figure~\ref{fig_coarse_grained_n_s_comparison_slopes_and_collapse}a shows our
simulation results for $r=0.1$ and different values of $f/p$.  One can
clearly see that the slope becomes steeper with decreasing $f/p$ and
appears to approach a limit
slope. Figure~\ref{fig_coarse_grained_n_s_comparison_slopes_and_collapse}b shows the slopes $d
\log{n(s)}/ds$ as function of $s$, indicating that the predicted limit
value 1.45 is indeed correct. 

Figure~\ref{fig_coarse_grained_n_s_comparison_slopes_and_collapse}c shows a collapse of the
cutoff parts of the curves, giving $\lambda \simeq  1.1$, just as in
the original forest-fire model. We also
performed a moment analysis of the fire-size distribution, giving the
same result $\lambda\simeq 1.1$. 

\begin{figure}
\begin{center}
\includegraphics[angle=0, width=\textwidth]{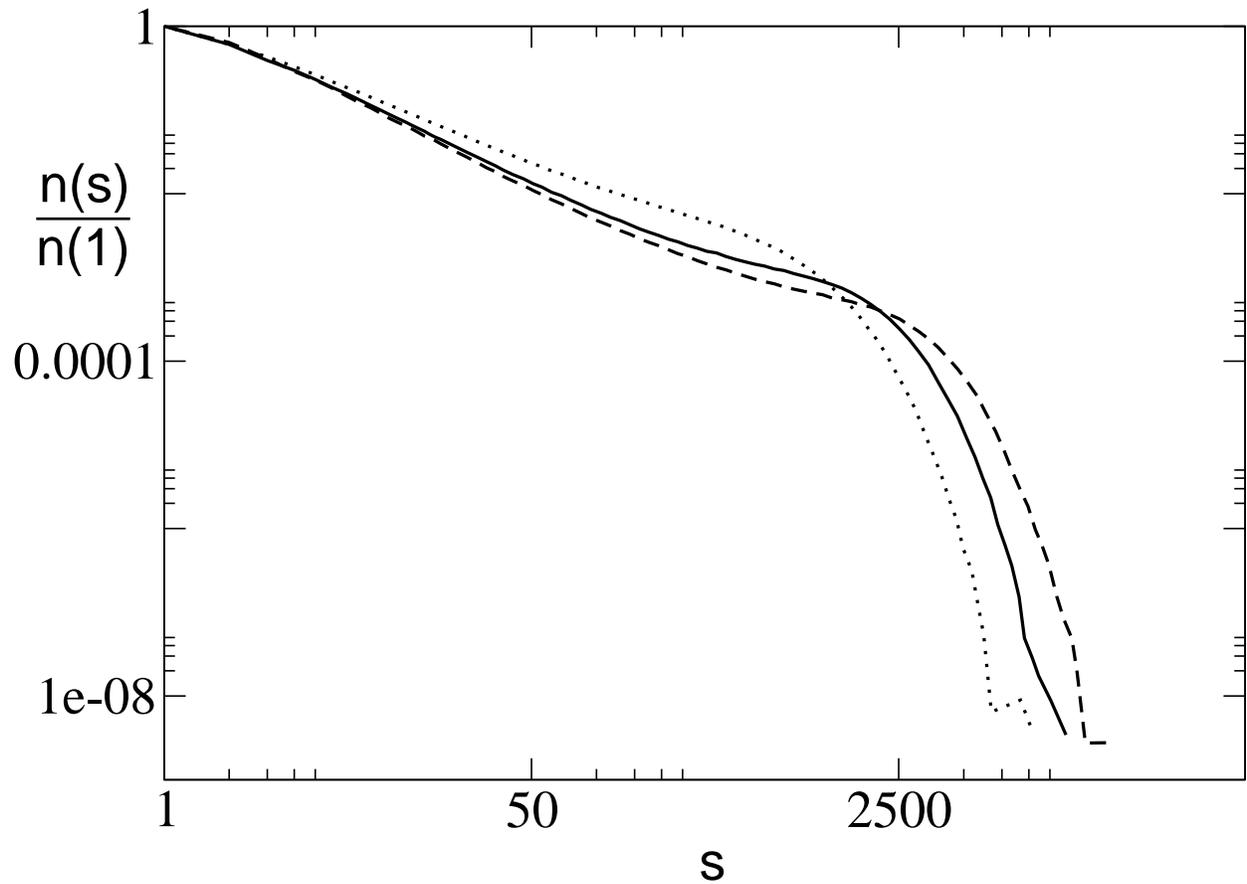}
\caption{Fire size distributions of the coarse grained model for
$r=0.1$ (dashed curve), $r=0.2$ (solid curve), and $r=0.5$ (dotted
curve).  $L=1000$ and $f/p$=0.01 for all systems.}
\label{fig_r_tendency}
\end{center}
\end{figure}

Figure~\ref{fig_r_tendency} shows the fire size distribution $n(s)$
for three different values of $r$, and for the same $f/p=0.01$.  For
smaller $r$, the slope becomes steeper and the cutoff bump becomes
more pronounced, indicating that for smaller $r$ the coarse-grained
model resembles the original model on larger scales. For smaller $r$,
the cutoff becomes larger. The reason is that for smaller $r$ a site
of the coarse-grained model corresponds to more sites of the original
model. The same lightning probability $f$ per site in the
coarse-grained model corresponds to a smaller lightning probability in
the original model when $r$ is smaller. 

Let us now estimate how many sites $z(r)$ of the original model
correspond to a site in the coarse-grained model with parameter $r$.
From Figure~\ref{fig_original_ffm_ns_superpos_and_tendency}a we find that
$$s_{\small\rm{max}} = A (f/p)^{-\lambda}$$
with $A \simeq 30$. Similarly, we have for the coarse-grained model
$$s_{\small\rm{max}} = B(r) (f/p)^{-\lambda} .$$ From
Figure~\ref{fig_coarse_grained_n_s_comparison_slopes_and_collapse}c we estimate $B(0.1)\simeq
100$, and from the data shown in Figure~\ref{fig_r_tendency} we then
obtain $B(0.2) \simeq 66$ and $B(0.5)\simeq 44$.  Now,
$s_{\small\rm{max}}$ sites in the coarse-grained model correspond to
$z(r)s_{\small\rm{max}}$ sites in the original model, and $f$ in the
coarse-grained model corresponds  to a lightning probability $f/z(r)$
per site in the original model. Therefore we have
$$B(r) (f/p)^{-\lambda}=A (f/pz)^{-\lambda}/z\, ,$$ leading to
$$z(r)=(B(r)/A)^{1/(\lambda-1)} \simeq (B(r)/A)^{10},$$
resulting in
$z(0.5)\simeq 46$, $z(0.2)\simeq 2650$, and $z(0.1)\simeq 170000$.
The length scales of the coarse-grained model are reduced by factors
of the order 7, 50, and 400 for these three $r$ values, compared to
the original model.

\begin{figure}
\begin{center}
\includegraphics[angle=0, width=\textwidth]{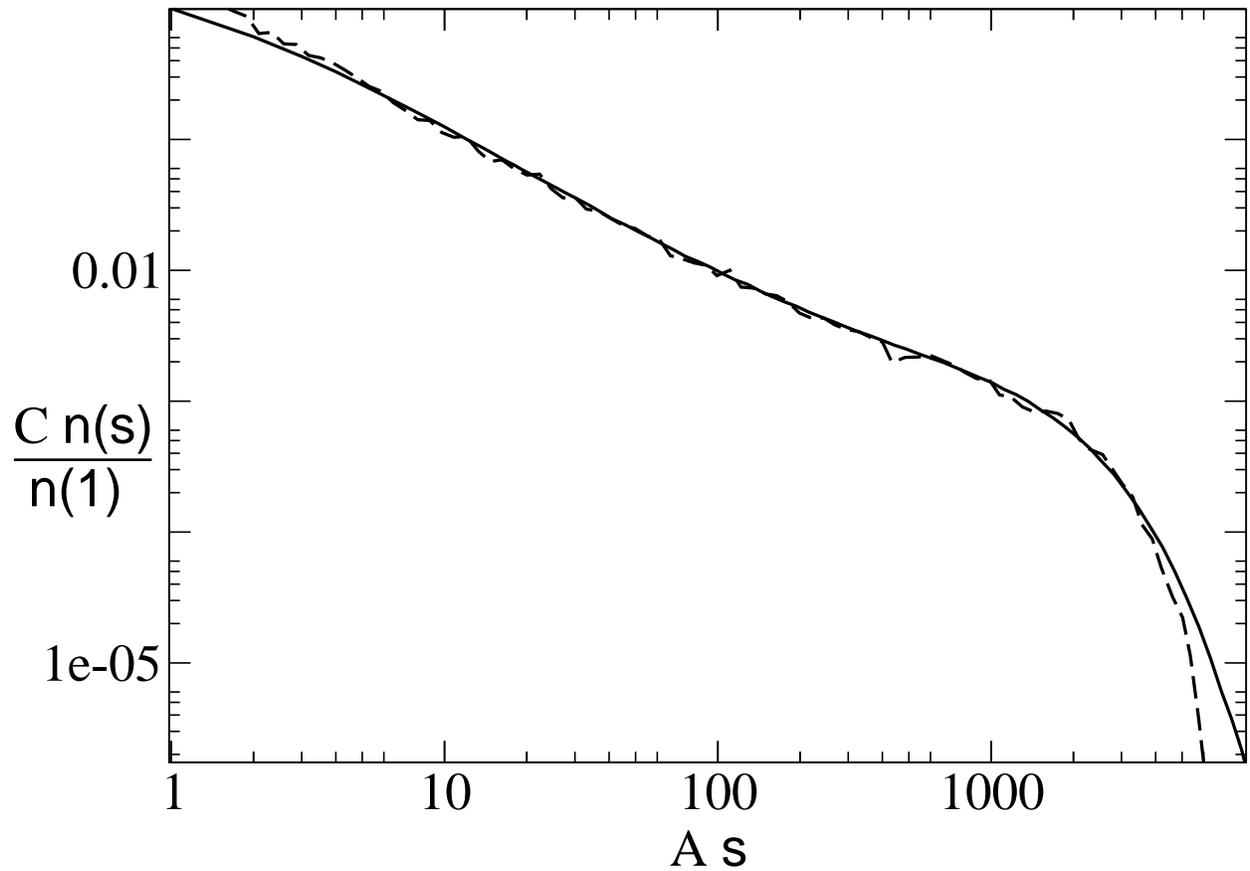}
\caption{Comparison of the fire distributions $n(s)$ 
  obtained for the original SOC forest fire model for $f=0.00000156$,
  $p=0.01$, and $L=800$ (dashed line, $C=41$, $A=1$), and for the
  coarse grained model for $r=0.5$, $f=0.00005$, $p=0.01$, and
  $L=1000$ (solid line, $C=1$, $A=44.9^{-1}$).  the ratio between the
  two cutoffs, $A$, is the same as the inverse ratio between the two
  lightning probabilities $f$. }
\label{fig_coarse_grained_n_s_comparison_with_non_coarse_grained}
\end{center}
\end{figure}

A direct comparison of a fire size distribution of the original model
and one of the coarse-grained model with $r=0.5$ confirms these
findings. We searched for two fire size distributions such that the
ratios of their $f$ values and the ratios of their
$s_{\small\rm{max}}$ values are similar. This ratio turned out to be
around 45, as shown in
Figure~\ref{fig_coarse_grained_n_s_comparison_with_non_coarse_grained},
and in agreement with the finding of the previous paragraph.
Figure~\ref{fig_coarse_grained_n_s_comparison_with_non_coarse_grained},
shows also that the shapes of the two fire-size distributions, while
similar, are not identical. Identical shapes cannot be expected, since
the coarse-grained model is not completely identical to the original
model on larger scales. For instance, inhomogeneities arising in the
original model within an area of size $z(r)$, cannot occur in the
coarse-grained model. This explains the difference in shape on small
scales.  On large scales, the difference in the shape of the cutoff is
probably due to the fact that the process of slow destruction of large
patches is slightly different in the two models. In both cases,
lightning strokes hitting the patch when its density is only slightly
above the percolation threshold make the patch more inhomogeneous.  In
the coarse-grained model, this leads to sites belonging to two widely
different density intervals, as mentioned further above in the context
of Figure~\ref{fig_snapshot_effective_model}. In the original model,
this leads to a couple of smaller dense tree clusters being left behind by
the fire, as can be concluded from Figure~\ref{fig_snapshot}.

To conclude this section, our coarse-grained model, while not being
exactly equivalent to the original model, shows the features expected
for the original model on larger scales and confirms in particular the
universality of the exponent $\lambda$ and our conjecture that the
exponent $\tau$ has an asymptotic value around 1.45.  The detailed
mechanism of birth and destruction of patches is somewhat different in
the two models and leads to different shapes of the fire size
distributions at small $s$ and for the largest $s$.

\section{Discussion}
\label{chap_discussion}

In this paper, we have argued that the fire size distribution in the
SOC FFM is the result of the superposition of two types of fires. The
smaller ones are fractal percolation clusters, while the larger fires
are compact and burn down a patch of a tree density above the
percolation threshold. We supported this picture by a direct analysis
of the model, by the artificial superposition of the two types of
fires, and by the introduction of a coarse-grained FFM. 

One of our main results is that the asymptotic exponent for the fire
size distribution is $\tau\simeq 1.45$, and is visible only at length
scales not accessible to present-day computer simulations.  For values
of the correlation length typically seen in computer simulations, the
exponent $\tau$ has an apparent value which is smaller, and which
seems to be insensitive to the lattice type used in the simulations.
Furthermore, we found that the cutoff exponent $\lambda$ has an
universal value $\lambda\simeq 1.1$, which is measured in the original
FFM as well as in the coarse-grained FFM for different simulation
parameters. The robustness of this exponent is additionally supported
by our earlier finding that the correlation length shows nice scaling
behavior in a generalized model where trees can be immune to fire
\cite{dro94}. 

We could not find the precise form of the size
distribution of patches, although we presented evidence that it should
be characterized by an exponent $b$ smaller than 1. The patch size
distribution is the result of a slow and highly nontrivial process of
birth and merging and destruction of patches. This process also
determines the size distribution of the large fires, for which we
could not give an analytical expression.

In contrast to the exponent $\tau$, we could not derive the value of
the exponent $\lambda$ from analytical arguments. We  cannot rule
out that its true value is $\lambda=1$, and that there are logarithmic
corrections which make it appear slightly larger than 1.

As we have shown, several scaling relations familiar from conventional
critical systems and in particular from percolation theory do not hold
in the SOC FFM.  Instead, the FFM is a new type of non equilibrium
critical system that has no equivalent in equilibrium physics. It is
characterized by different phenomena on different scales, and by a
patchy structure indicating that neighboring sites tend to be
synchronized by burning down during the same large fires. 

By introducing the coarse-grained model, we have shown that there
exists an entire class of models that share the same main features of
a patchy structure and two qualitatively different types of fires, the
asymptotic exponent $\tau\simeq 1.45$, and the cutoff exponent $\lambda
\simeq 1.1$, while details like the precise shape of the cutoff and
the precise mechanism of birth and destruction of patches may differ.

Our results show that SOC in dissipative systems can be caused by
mechanisms fundamentally different from equilibrium critical
phenomena. We expect that other dissipative SOC systems are driven to
criticality by mechanisms similar to the ones found in the FFM.  This
applies in particular to the SOC earthquake model \cite{ola92}, where
a patchy structure with partial synchronization of neighboring sites
was also found \cite{mid95}. Very recently, it was also found that
this model contains two qualitatively different types of avalanches:
those within a ``patch'', and those that enter it from outside and
span the entire ``patch'' \cite{lis01}.

\acknowledgements
This work was supported by the Deutsche 
Forschungsgemeinschaft (DFG) under Contract
No Dr300-2/1, and by the EU network project (TMR) ``fractal
structures and self organization'', EU-contract ERPFMRXCT980183.

% \end{multicols}


\begin{references}
\bibitem{bak87}  P. Bak, C. Tang and K. Wiesenfeld,
                 Phys. Rev. Lett. {\bf 59}, 381 (1987);
                 Phys. Rev. A {\bf 38}, 364 (1988);
                 P. Bak and M. Creutz, in {\it Fractals in Science},
                 ed.\ by A. Bunde and S. Havlin, (Springer, Berlin, 1994).
\bibitem{dro92}  B. Drossel and F. Schwabl,
                 Phys. Rev. Lett. {\bf 69}, 1629 (1992).
\bibitem{hen93}  C.L. Henley, Phys. Rev. Lett. {\bf 71}, 2741 (1993).
\bibitem{gra93}  P. Grassberger,
                 J. Phys. A {\bf 26}, 2081 (1993).
\bibitem{cla94}  S. Clar, B. Drossel, and F. Schwabl,
                 Phys. Rev. E {\bf 50}, 1009 (1994).
\bibitem{ola92}  Z. Olami, H. J. S. Feder and K. Christensen,
                 Phys. Rev. Lett. {\bf 68}, 1244 (1992).
\bibitem{sne93}  K. Sneppen, and P. Bak,
                 Phys. Rev. Lett. {\bf 71}, 4083 (1993);
\bibitem{PMB}    M. Paczuski, S. Maslov, and P. Bak, Phys. Rev. E {\bf 53},
                 414 (1996).
\bibitem{teb98}  M. De Menech, A. L. Stella, C. Tebaldi,
                 Phys. Rev. E {\bf 58}, R2677 (1998).
\bibitem{ste99}  C. Tebaldi, M. De Menech, and A.L. Stella,
                 Phys. Rev. Lett. {\bf 83}, 3952 (1999). 
\bibitem{dro00}  B. Drossel, Phys. Rev. E {\bf 61}, R2168 (2000). 
\bibitem{pac01}  S. Lise and M. Paczuski, Phys. Rev. E {\bf 63}, 036111 (2001).
\bibitem{hon96}  A. Honecker and I. Peschel,
                 Physica A {\bf 239}, 509 (1997).
\bibitem{cla95}  S. Clar, B. Drossel, and F. Schwabl,
                 Phys. Rev. Lett. {\bf 75}, 2722 (1995); 
                 S. Clar, K. Schenk, and F. Schwabl,
                 Phys. Rev. E {\bf 55}, 2174 (1997).
\bibitem{cla95b} S. Clar, B. Drossel, K. Schenk and F. Schwabl,
                 Phys. Rev. E {\bf 56}, 2467 (1997); 
\bibitem{sch00}  K. Schenk, B. Drossel, S. Clar, F. Schwabl, 
                 Eur. Phys. J. B {\bf 15}, 177 (2000). 
\bibitem{sta92} D. Stauffer and A. Aharony, 
                {\em Introduction to Percolation Theory}, 
                (Taylor and Francis, London, 1992).  
\bibitem{pas00} R. Pastor-Satorras and A. Vespignani, 
                Phys. Rev. E {\bf 61}, 4854 (2000). 
\bibitem{dro94} B. Drossel, S. Clar, and F. Schwabl, 
                Phys. Rev. E {\bf 50}, R2399 (1994).
\bibitem{mid95} A.A. Middleton and C. Tang, 
                Phys. Rev. Lett. {\bf 74}, 742 (1995).
\bibitem{lis01} S. Lise and M. Paczuski, cond-mat/0104032.
\end{references}
\end{document}